\documentclass[%
 reprint,
 amsmath,amssymb,
 aps,
]{revtex4-1}

\usepackage{graphicx}
\usepackage{dcolumn}
\usepackage{bm}
\usepackage{hyperref}
\usepackage[mathlines]{lineno}
\usepackage{color}
\usepackage[dvipsnames]{xcolor}
\usepackage{float}
\usepackage{soul}

\begin{document}

\preprint{WTF-draft}

\title{Curvature controlled pattern formation in floating shells
}

\author{Octavio Albarr\'an}
\email{octavio.albarran@ds.mpg.de}
\affiliation{%
Max Planck Institute for Dynamics and Self-Organization,
Am Fassberg 17, 37077 G\"ottingen, Germany
 }%
\author{Desislava V. Todorova}
\email{tdes@sas.upenn.edu}
\author{Eleni Katifori}%
\email{katifori@sas.upenn.edu}
\affiliation{%
 Department of Physics and Astronomy, University of Pennsylvania, Philadelphia, Pennsylvania 19104, USA
}%
\author{Lucas Goehring}%
\email{lucas.goehring@ntu.ac.uk}
\affiliation{%
    School of Science and Technology, Nottingham Trent University, Nottingham NG11 8NS, UK
}%

             
\begin{abstract}
Shells, when confined, can deform in a broad assortment of shapes and patterns, often quite dissimilar to what is produced by their flat counterparts (plates). In this work we discuss the morphological landscape of shells deposited on a fluid substrate. Floating shells spontaneously buckle to accommodate the natural excess of projected area and, depending on their intrinsic properties, structured wrinkling configurations emerge. We examine the mechanics of these instabilities and provide a theoretical framework to link the geometry of the shell with a space-dependent confinement. Finally, we discuss the potential of harnessing geometry and intrinsic curvature as new tools for controlled fabrication of patterns on thin surfaces.  
\end{abstract}


\maketitle
%
%


Mechanical instabilities of confined thin sheets, films or membranes have been investigated extensively for both their relevance to biological systems and their importance in engineering.  Applications include the fabrication of flexible microelectronics and microfluidic devices~\cite{Genzer2006, Ohzono2009, Yang_review2010, Rogers2010, Cui2016}, complex ordered designs~\cite{Ding2013, Chen2013, Zhang2017}, biological templates~\cite{jiang2002controlling}, electronic skin~\cite{wagner2004electronic, hammock201325th}, optical devices~\cite{harrison2004sinusoidal}, metrology standards~\cite{Schweikart2009}, and so on. In a typical set-up a thin sheet is deposited onto a softer, or fluid, substrate~\cite{Brau2013}. This film is then compressed, either by external forces or by a differential growth between the film and substrate~\cite{Audoly2011, pineirua2013capillary, Paulsen2016, bense2017buckling}. When the stresses are above a certain threshold, the flat film becomes unstable.  Depending on the system, a series of different processes can then take place, including spontaneous wrinkling, the continuous transition from wrinkles to localized folds, wrinkle period doubling,  hierarchical buckling and folding~\cite{Kim2011, Leahy2010, huntington2014controlling, gemmer2016isometric}, the growth of herringbone and labyrinth-like structures~\cite{Lin2007, Efimenko2005, Vandeparre2008}, and wrinkle-induced delamination~\cite{Nolte2017}. Similar types of complex phenomena have been reported for both flat sheets and curved interfaces~\cite{King2012, Yao2013, Aharoni2017}.

Generally, however, most detailed studies have considered the reference configuration of the deposited film to be flat.   Confinement is then presented as a uniform property of the foundation, generated by a confining pressure or misfit stress, for example. In contrast, when the natural curvature of the sheet is non-zero the system must satisfy geometric constraints leading to a local concept of confinement, which can now vary dramatically across space. In this case the mechanical responses should be treated in the context of shell mechanics.  

Shells are fundamentally different from plates because of their geometric rigidity~\cite{landau1986theory,ElasticityGeometryAudoly}.  This leads, for instance, to the classic analogy between the symmetrical buckling of cylindrical shells under axial compression and the wrinkling of sheets on elastic foundations \cite{timoshenko1961theory}. Furthermore, in shells bending and stretching are intimately related. This coupling stabilizes localized deformations and can drive the spontaneous emergence of wrinkles, folds, dimples and blisters, even without any restoring elastic substrate~\cite{pogorelov1988bendings, Efimenko2005, Ebrahimi2014}. 

In this work we present an experimental system for studying mechanical instabilities in shells with strong natural curvature, and provide a theoretical framework capable of explaining the origins and scaling of the instabilities found therein. In particular, we consider a soft shell, approximately 5 cm in size and about 30 $\mu$m in thickness, deposited on a fluid like water. The shell remains at the fluid surface by buoyancy, adopting a geometry that we will call a \textit{floating shell}, in analogy to the floating elastica~\cite{Audoly2011}. The shell's final configuration is driven by a hydrostatic flattening effect: the specific weight of the liquid mimics the stiffness of a flat elastic substrate~\cite{Pocivavsek2008}.  As such, much of the results presented here will also apply to shells on solid foundations with incommensurate curvatures. The system we study is self-organized in the sense that its lateral confinement is an intrinsic property appearing from geometrical frustration and does not require external driving forces. It offers a platform to study the consequences of bending-stretching coupling, and  mechanisms to harness the arising instabilities for controlled pattern formation.

\begin{figure*}[ht]
\centering
\includegraphics[width=\linewidth]{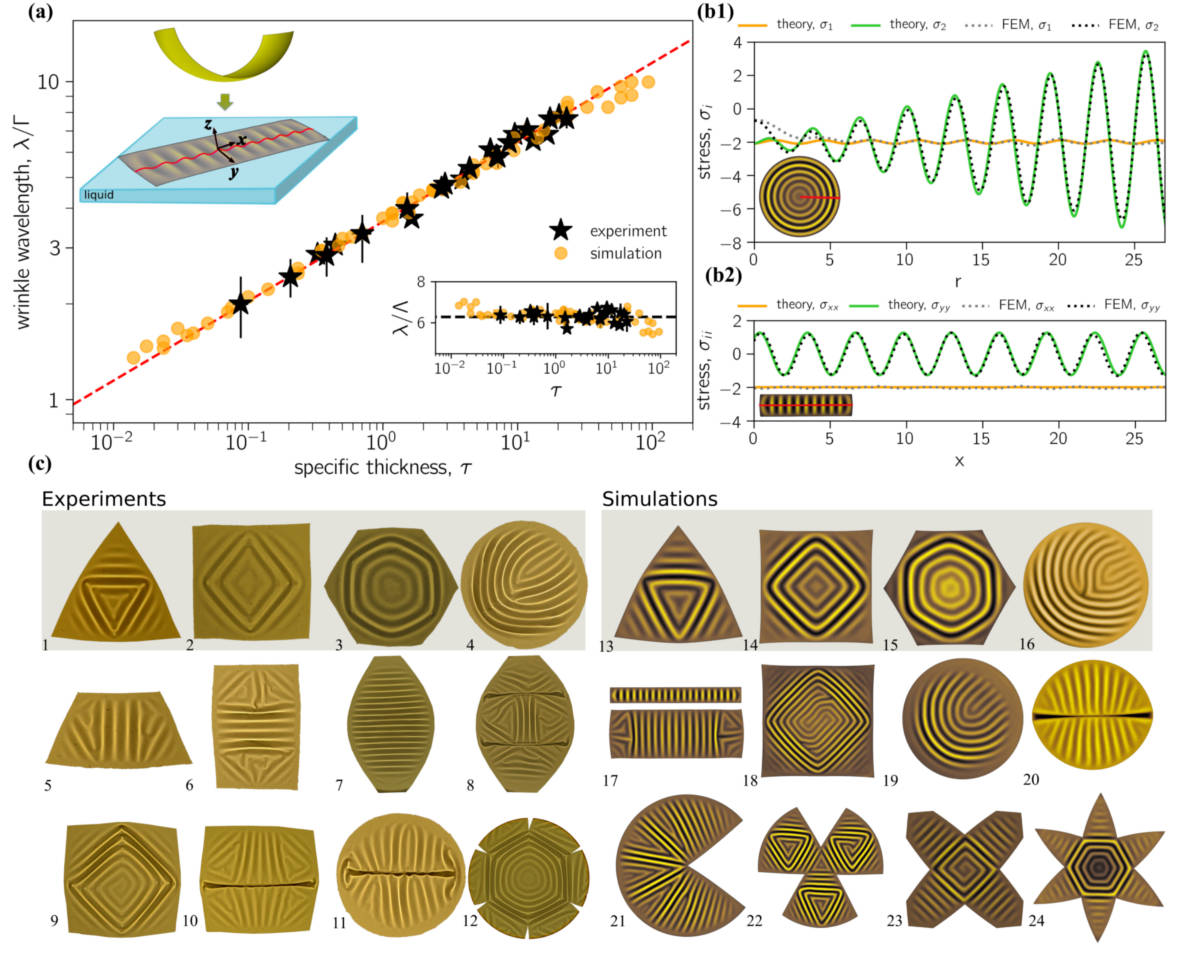}
\caption{Wrinkling configurations in floating shells. (a) The wrinkle wavelength, $\lambda$, normalized by the deformation length scale $\Gamma=\sqrt{Rt}$ (main plot) and $\Lambda=\left( K_B / K_g \right)^{1/4}$ (inset), versus the specific thickness of the shell, $\tau = {Et}/{R^2 K_g}$. Representative error bars are shown on a subset of the experimental results. The dashed lines indicate the elastica model for the different normalizations, $\lambda/\Gamma = 2\pi(\tau/9)^{1/4}$ (red dashed line) and $\lambda/\Lambda = 2\pi$ (black dashed line). The upper inset shows a schematic of the floating shell setup. (b) The in-plane principal stress profiles (along the red lines drawn in the insets) in: (b1) an axisymmetric configuration of a spherical cup ($\tau = 0.85$, $R = 6$ cm) and (b2) an anisotropic configuration of a spherical strip ($\tau = 5.0$,  $R = 3.8$ cm). The theoretical trends are given by Eqs.~\ref{StressResultAxissymm} and \ref{AnisotropicStressSolution}.  (c) A constellation of patterns in floating shells of various geometries are shown, drawn from (left) experiments and (right) simulations, on the range $0.1 < \tau < 20$.  Experimentally, shells have been imaged by a shadowgraph (see Methods); simulation images depict surface displacement, using a similar colour scheme. We can observe patterns comprising single (7, 17) and multiple (\textit{e.g.} 1-3) domains of straight wrinkles (7, 17, 21), curved wrinkles (4,16,19), axisymmetric wrinkles (b1, inset), folds (8, 10, 11, 20), flattened boundaries (19), dislocations (4, 16) amongst other configurations.  The top row compares experimental and simulated patterns for similar geometries and material properties. 
}
\label{fig:Fig1}
\end{figure*}

\section*{Floating Shells} 

To probe how intrinsic curvature affects pattern formation driven by elastic instabilities, we combine experimental investigation with the finite element simulation (FEM) of thin curved shells on a liquid substrate (see Methods). Shells of various shapes, sizes and thicknesses are cut out of hemispheres, and thus have a constant Gaussian curvature.  When floated on a liquid surface the shells assume a stunning variety of patterns by harvesting diverse instabilities (see Fig.\ref{fig:Fig1}).  The patterns typically show well-defined wrinkles of constant wavelength, $\lambda$.  The scaling of this wavelength is demonstrated in Fig.~\ref{fig:Fig1}(a). Depending on their configuration, they may also exhibit a spatially varying amplitude, as reflected by variations in colour intensity of the top-view images provided in Fig.\ref{fig:Fig1}(b,c).  

Apart from their wavelength and amplitude, the topological characteristics of wrinkles can be described in terms of ridge lines.  These continuous curves trace along the set of locally maximal points that define the crest of each wrinkle, and appear as bright lines in the images shown in Fig.\ref{fig:Fig1}(c). When seen from above, as there, these lines can be straight, smoothly curved or sharply bent.  In this last case the ridge line changes its direction on a length-scale comparable to $\lambda$. The length of a ridge line between localised bends can be used to define the size, $D$, of a wrinkle domain. 

The agreement between our simulations and experiments is demonstrated in Fig.~\ref{fig:Fig1}(c), the first row of which compares four paired sets of floating shell patterns. Additional examples show the variety of stable configurations that are possible, including distinct morphological domains of parallel, straight wrinkles, domain boundaries, defects such as dislocations, localised features such as folds, and boundary effects such as the damping of wrinkles around flattened edges. 

The particular configuration of any wrinkle pattern is strongly influenced by the shape of its boundary. The edges of the floating shell impose traction-free boundary conditions, such that the in-plane normal stresses there are zero. Thus, in the extreme case of shells cut into strips or slices, with two long edges, wrinkles will tend to align perpendicular to the boundaries, resulting in a single domain of straight wrinkles. This is demonstrated in Fig.~\ref{fig:Fig1}(c.17), and is a reference configuration that we will study here in depth. Generally, gently curved edges in high aspect ratio samples can still yield single straight wrinkle domains, as in Fig.~\ref{fig:Fig1}(c.7). 

Multiple domains appear in samples with small aspect ratios where at least three edges intervene (see Fig.\ref{fig:Fig1}(c.1,c.2)). The symmetry of the shell influences the number of domains that may be found in the final pattern, \textit{e.g.} three in a triangle, four in a square, \textit{etc}.  These domains are separated by grain boundaries or folds. Recently, we used the floating shell setup to study some of the characteristic features of these domains, including the typical boundary width and characteristic domain size, using an approach based on the theory of smectic liquid crystals~\cite{Aharoni2017}. 

Bent wrinkles emerge in samples with many edges or highly curved edges (see Fig.\ref{fig:Fig1}(c.3,c.4,c.12)). Axisymmetric solutions are also achievable, as in Fig.~\ref{fig:Fig1}(b.1), and we will discuss them more in depth in the next section.  

Finally, we find that patterns are controlled by the amount of excess area in the shell, \textit{i.e.} the difference between the actual surface area of the undeformed shell, and its projected area when floating.  For large excess areas, wrinkle patterns may become metastable and allow for different configurations. Localized structures also absorb excess material in the most efficient way, so, as confinement is increased, grain boundaries can form spontaneously and evolve into folds. A wrinkle-to-fold transition can be seen in Fig.\ref{fig:Fig1} between panels 7 and 8, and more details of this transition are presented in Fig.~\ref{fig:WTF-slices}. There we show that a small change in the size of the sample prompts a transition from single to multiple domains. We characterise this process and discuss the conditions that drive domain reconfiguration.

\section*{Analysis} 

In addition to the shape of its boundary, a floating shell is controlled by four mechanical parameters. These are its in-plane stretching modulus $Y_S$, bending stiffness $K_B$, radius of curvature $R$ and the specific weight of its supporting fluid, $K_g\equiv \rho g$, where $\rho$ is the fluid's density and $g$ the acceleration due to gravity.  In our experiments any surface tension is orders of magnitude lower than the stretching modulus, and does not significantly affect the patterns formed~\cite{Aharoni2017}. We thus neglect surface tension in our simulations and model. The shells are considered to be composed of linear elastic materials with homogeneous thickness $t$, Young's modulus $E$ and Poisson's ratio $\nu=1/2$, so that $Y_S = 4 E t / 3$ and $K_B = E t^3/9$. 

Dimensional analysis of the above system implies that its equilibrium equations depend on two dimensionless groups, at most.  These groups can be taken to be ratios of three independent length scales of deformation.  One of these lengths is the shell radius, $R$. Analogous to floating elasticas~\cite{Pocivavsek2008, Diamant2011}, a second natural length of the form $\Lambda = ({K_B}/{K_g})^{1/4}$ arises from balancing the bending energy, $U_B \sim K_B {\Lambda^2}/{R^2}$, of a dimple of size $\Lambda$~\cite{pogorelov1988bendings} to its gravitational energy, $U_g \sim K_g {\Lambda^6}/{R^2}$. For a shell, a third length, $\Gamma = \sqrt{R t }$, represents the scale at which the bending and stretching energies ($U_S \sim Y_S {\Gamma^6}/{R^4}$) required to flatten the shell are comparable.  This scale is also important for finding the critical conditions for buckling \cite{pauchard1998contact} and the wavelength of anisotropic blisters \cite{Ebrahimi2014}. It is a fundamental length of the elastic theory of shells and is usually introduced as the size of a region over which a shell reacts against localized forces \cite{landau1986theory}.

The regime which we explore here is that of a strong liquid interaction, $\Lambda \ll R $, and the thin shell approximation, $\Gamma \ll R$. Due to this scale separation the normalized wavelength of the patterns (\textit{i.e.} either $\lambda/ \Gamma$ or $\lambda/ \Lambda$) is expected to depend on only one dimensionless parameter, at most.  This is taken to be the specific thickness $\tau \equiv {Et}/{R^2 K_g} \sim \left({\Lambda}/{\Gamma}\right)^4$, related to the \textit{bendability} of the shell~\cite{Davidovitch2011, King2012}. In fact, $\tau \sim {K_B}/{K_g \Gamma^4}$ is also an effective bending rigidity and, together with the normalized natural curvature $\kappa_0 \equiv {\Lambda}/{R}$, settles the dimensionless groups used in this work. 

Figure~\ref{fig:Fig1}(a) shows results for the relative wavelengths $\lambda / \Gamma$ (and $\lambda / \Lambda$ in the inset plot) as measured over a wide range of specific thicknesses, $\tau$. We include data from both experiments and finite element simulations, which are in excellent agreement with each other. The dashed lines indicate the elastica model for the different normalizations, $\lambda/\Gamma = 2\pi(\tau/9)^{1/4}$ (red dashed line) and $\lambda/\Lambda = 2\pi$ (black dashed line). In Fig.~\ref{fig:vart} we demonstrate this scaling visually by varying $\tau$ on shells of an otherwise fixed geometry.

In the next sections we will rationalize the general features observed in floating shell patterns by studying two idealized scenarios: axisymmetric and purely anisotropic configurations. Hereafter, we also normalize all lengths by $\Lambda$ and energy densities (per unit area) by $K_g \Lambda^2$.


{\bf\textit{Axisymmetric configurations.}} 
For small local vertical displacement, $h$, measured from the free surface of the liquid, the dimensionless energy density of an axisymmetric configuration of a shallow shell is (Eq.~\ref{DimensionLessTotalEnergy3}, derived in the supplemental information),
\begin{equation}
\begin{split}
\mathcal{F} & = \frac{1}{2} \left[  \ddot{h}^2 + \left( \frac{\dot{h}}{r} \right)^2 - 2 \kappa_0 \left( \ddot{h} + \frac{\dot{h}}{r} \right) \right]  +  \frac{1}{2} h^2 + \\ & + \frac{\eta}{2} \left( \epsilon_1^{\, 2} + \epsilon_2^{\, 2} + \epsilon_1 \epsilon_2 \right)  \text{,}
\end{split}
\label{DimensionLessTotalEnergy} 
\end{equation} 
where $r$ is the radial coordinate and the over-dots denote $r$-derivatives. The terms in the square brackets account for the bending energy, while $\frac{1}{2} h^2$ represents the gravitational potential. The last term in Eq.\ref{DimensionLessTotalEnergy} is the stretching energy where $\eta = \frac{Y_S}{K_g \Lambda^2} = \frac{4}{3} \frac{\tau}{\kappa_0^{\ 2}}$ is the stretching rigidity and $\epsilon_{1}$, $\epsilon_{2}$ are the meridional and hoop strains. For small deformations, the strains are coupled to the local deflections through $\varepsilon \equiv \epsilon_2 + r \dot{\epsilon_2} - \epsilon_1 + \frac{1}{2} \dot{h}^2 =  \frac{1}{2} \kappa_0^{\ 2} r^2 $, where the function $\varepsilon(r,\epsilon_1, \epsilon_2, \dot{\epsilon_2}, \dot{h}) = \frac{1}{2} \kappa_0^{\ 2} r^2 \ll 1 $ defines the lateral excess of length imposed by the shell (Eq.~\ref{DeflectionStretching2}).

Unlike the floating \textit{elastica}, where stretching may be discounted, here all contributions to the energy density need to be taken into consideration. For instance, whenever wrinkling patterns emerge, the local amplitude becomes proportional to $\sqrt{\varepsilon}$, implying that both the bending energy and the gravitational potential scale linearly with $\varepsilon$. On the other hand, the magnitude of the stresses in a wrinkling pattern is at least of order one, meaning that the strains scale with $\sim 1/\eta $. Comparison between $\varepsilon \sim \kappa_0^{\ 2}$ and $1/\eta \sim \kappa_0^{\ 2}/ \tau $ indicates that stretching becomes increasingly dominant for smaller $\tau$. We will show that when $\tau \gg 1$ the maximum stresses are proportional to $\tau$. Thus, the fundamental aspects of wrinkling in shells cannot be captured without introducing local stretching. 

Stable configurations minimise the surface integral of Eq.~\ref{DimensionLessTotalEnergy}. The resulting equilibrium equations are derived in the supplemental information as (Eqs.~\ref{StressBalance2},\ref{DeflectionEquation2})
\begin{equation}
r^2 \ddot{\sigma_1} + 3r \dot{\sigma_1} = \frac{3 \eta }{8} \left(  2 \varepsilon -  \dot{h}^2 \right)  
\label{StressEquation} \text{,}
\end{equation}
\begin{equation}
\ddddot{h} - \sigma_1 \ddot{h} + h + \frac{2\dddot{h} - \sigma_1 \dot{h}}{r} =  \frac{\ddot{h}}{r^2}  + \left( \dot{\sigma_1} - \frac{1}{r^3} \right) \dot{h}  
\label{DeflectionEquation} \text{\ ,}
\end{equation}
where $\sigma_1=\eta \left( \epsilon_1 + \frac{1}{2}\epsilon_2 \right)$ is the meridional stress. Eq.~\ref{DeflectionEquation} has the same structure as the first F{\"o}ppl-von K\'arm\'an equation (Eq.~\ref{vonKarmanEquations}), implying that the curved shell behaves as a flat disc with residual meridional stress $\widetilde{\sigma}_1 = \frac{3 \eta}{32} \kappa_0^{\ 2} r^2$. If $T_1$ is the radial stress satisfying the equilibrium equations on a disc, we can decompose the stress of the shell as $\sigma_1=\widetilde{\sigma}_{1} + T_1$, thus formulating the mechanics from the point of view of the metric on a plane. From this perspective, the effect of the natural curvature is similar to that of a confining pressure along the shell boundary.

Equations \ref{StressEquation} and \ref{DeflectionEquation} can be solved using a perturbation scheme with $\kappa_0$ as the small parameter. Although the system of equations is derived for thin shells and small deflections, the second order solutions retain the fundamental features of buckling and wrinkling instabilities, and show exceptional agreement with observations.
  
To order \textit{zero}, Eq.~\ref{StressEquation} yields $ \sigma_1 = \sigma_1^{(0)} + \frac{c_1}{r^2} $, where $\sigma_1^{(0)}$ and $c_1$ are constants. To first order, Eq.~\ref{DeflectionEquation} has sinusoidal solutions $h= A \cos( q r + \phi)$ for the unique combination $\sigma_1^{(0)}=-2$, $c_1=-1$, and $q=1$. The corresponding dimensionless wavelength is equal to the value obtained for the floating elastica, $\lambda / \Lambda = 2 \pi$. In the limit $r \ll 1$, wrinkling solutions are not appropriate and we need to set $c_1=0$, as otherwise the meridional stress becomes unbounded. In this situation the terms proportional to the negative powers of $r$ in Eq.~\ref{DeflectionEquation} become dominant, thus $r \ddot{h} - \dot{h} \approx 0 $, suggesting a \textit{dimple}-like boundary layer of $h(r)= \frac{1}{2} \kappa_0 r^2$, where the natural geometry of the shell is conserved.  We shall see again similar localized structures in the context of straight wrinkles.

Because of the symmetry $h(\kappa_0) = - h(-\kappa_0)$, the harmonic wrinkling solution is valid up to second order perturbations, and it holds for any slowly varying amplitude envelope, such that $\dot{A}(r) \ll A(r)$. We can thus determine the specific form of $A(r)$ by solving Eq.~\ref{StressEquation} to higher order. 

Since $\dot{h}^2= \frac{A^2}{2} \left[ 1 - \cos 2(r+\phi ) \right]$, a stable wrinkling pattern requires that $A = 2 \sqrt{\varepsilon} =\sqrt{2}\kappa_0 r$, in order for the average value of the right side of Eq.~\ref{StressEquation} to be zero, and the stresses minimised. Thus, a general feature of wrinkles emerges: the longitudinal midlines (red lines drawn, for instance, in the schematic of Fig.~\ref{fig:Fig1}(a), Fig.~\ref{fig:Fig1}(b)  and Fig.~\ref{sketch2}) have approximately the same arc-length as in their undeformed state, meaning that the excess length follows the classic relationship $\varepsilon = \frac{1}{4} q^2 A^2 $ \cite{cerda2002thin, Mahadevan2003}.  This result also holds for the amplitude profiles in non-axisymmetric geometries. For instance, the local excess length of a straight wrinkle of dimensionless length $D \gg 1$ is $\varepsilon = \frac{1}{8} \kappa_0^{\ 2} D^2$, and $ A=\frac{\kappa_0}{\sqrt{2}} \frac{D}{q}$ is the amplitude of the associated sinusoidal profile.

To second order in $\kappa_0$, Eq.~\ref{StressEquation} reads $\ddot{\sigma_1} + 3\frac{\dot{\sigma_1}}{r} = \frac{\tau}{2} \cos 2(r+\phi) $. Since the stresses satisfy the stress equilibrium $ \sigma_2=\sigma_1 + r \dot{\sigma_1} $, for $r \gg 1$ we find that
\begin{equation}
\begin{split}
\sigma_1 & = -2 -\frac{\tau}{8} \cos 2(r+\phi) \text{ ,}\\
\sigma_2 & = -2 + \frac{\tau}{4} \left[ r \sin 2(r+\phi) - \frac{1}{2} \cos 2(r+\phi) \right] \text{.}  
\end{split}
\label{StressResultAxissymm} 
\end{equation}
These expressions illuminate how the geometric rigidity of the shells determines their stress distribution. The stress oscillations, of period $\lambda/2$, are significant. In Fig.~\ref{fig:Fig1}(b.1), we compare analytical solutions from Eq.~\ref{StressResultAxissymm} to measured stresses obtained from finite element simulations. The amplitude of the hoop stress, $\sigma_2$, grows linearly with the radial position, in contrast to a floating elastica, where the stresses, to second order, are $\sigma_1=-2$ and $\sigma_2=0$~\cite{Diamant2011}.  In Fig.~\ref{fig:Fig1}(b.2), we show typical stress profiles for straight wrinkles, which we describe next.

\begin{figure*}[htpb]
\centering
\includegraphics[width=\linewidth]{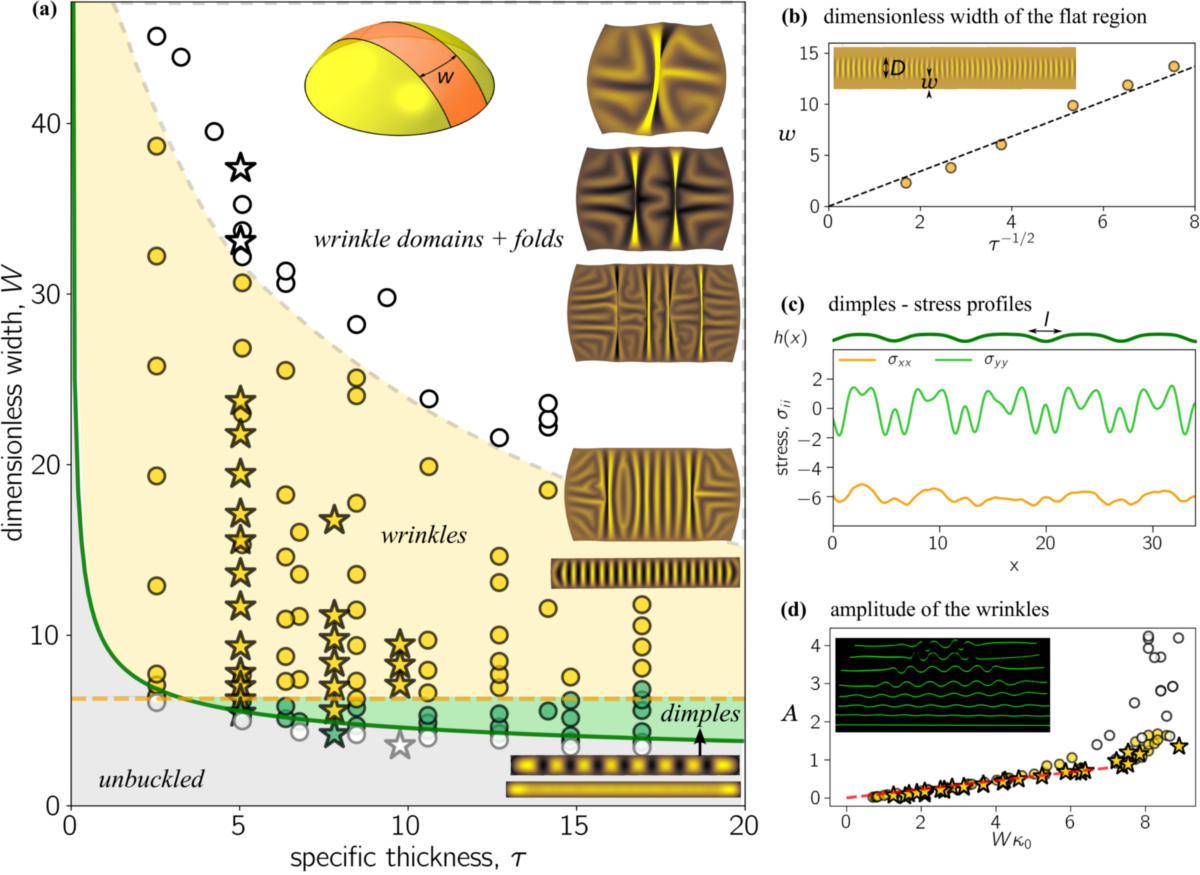}
\caption{Pattern selection in spherical strips. (a) The normalized strip width and shell thickness determine the type of pattern formed. Circles represent  simulations, and the stars, experiments. The colour-coding distinguishes between four types of patterns: unbuckled strips (grey), periodic dimples (green), straight wrinkles (yellow), and multiple domains and/or folds (white). Representative configurations are shown on the right. The orange dashed line shows $W = \lambda = 2\pi \Lambda$, where $\lambda$ is the measured periodicity of the pattern. The green line corresponds to the numerical solution of Eq.~\ref{EqForWc}. The dashed grey line separates configurations with multiple domains from single-domain solutions. (b) The width of the flat region along the border of a floating shell, $w$, is controlled by its specific thickness, $\tau$.  Here, we compare the theoretical prediction of $w \approx {1.70}/{\sqrt{\tau}}$ (dashed line) with simulation results (circles). (c) An example is shown of the mid-line profile of a fragment of a strip with periodic dimples, along with the corresponding stresses along the central line measured from simulations. (d) The amplitude of straight wrinkles depends on the excess area needing to be accommodated, and hence grows proportionally to $\kappa_0 W$, until a critical point, when buckling occurs.  The red line corresponds to the predicted behaviour for small amplitudes, $A={\kappa_0 W}/{q \sqrt{2}}$. Inset: experimental snapshots of the mid-line profiles of floating strips, increasing in width from bottom to top.
}
\label{fig:PhaseDiagram}
\end{figure*}

{\bf\textit{Anisotropic configurations.}} 
Long spherical strips (rectangular cuts from a spherical cap) are the simplest platform to study anisotropic configurations, as the two long and equally-spaced edges dominate the pattern selection.  Experimentally, depending on the relative separation of the edges ($W$, the dimensionless projected width of the strip), the excess of area results in three distinct scenarios.  When $W \ll 1$, floating strips adopt unbuckled configurations of zero Gaussian curvature. For $W$ larger than a critical value, $W_c(\tau)$, strips buckle and exhibit a single patterned domain. Finally, when $W \gg W_c$, multiple domains coexist, separated by localised features. In Fig.~\ref{fig:PhaseDiagram}(a), experimental and numerical data are presented in a colour-coded phase diagram characterising those morphologies. Next, a systematic analysis of the observed instabilities is presented. A mathematical treatment of a more general case involving toroidal strips (rectangular cuts from toroidal caps) is provided in the Supplementary Information, \ref{sec:toroidal1} and \ref{sec:toroidal2}.

Our analysis is based on the concept of residual stresses. As above, to second order the vertical shell displacement, $h(x,y)$, satisfies the first F{\"o}ppl-von K\'arm\'an equation with stresses $\sigma_{ik} = \widetilde{\sigma}_{ik} + T_{ik}$, where $\widetilde{\sigma}_{ik}$ are the residual stress components due to a planar flattening of the shell and $T_{ik}$ satisfies the stress equilibrium of a plate.

{\it (1) Unbuckled (planar) configurations.} 
In a planar configuration, the curvatures along the long axis ($x-$axis) of the strip  vanish. The deformation can be understood by treating the strip as a naturally curved beam with a cross-sectional profile $z(y) \approx \frac{1}{2} \kappa_0 y^2$. Thus, when the strip is bent perpendicularly to and kept straight along the $x$-axis, the longitudinal stress is $\widetilde{\sigma}_{xx} = \frac{3 \eta}{4} \left[ \kappa_0 z(y) - \epsilon \right]$ (see \ref{sec:toroidal1}), where $\epsilon$ is the strain along the central line and $\widetilde{\sigma}_{yy}=\widetilde{\sigma}_{xy}=0$.  The minimization of the stretching energy, which is proportional to $\int_{-W/2}^{W/2} \widetilde{\sigma}_{xx}^{\ 2} dy$, yields $\epsilon=\frac{1}{24} \kappa_0^{\ 2} W^2 $. Then, $\widetilde{\sigma}_{xx}=\frac{\tau}{2} \left( y^2  - \frac{1}{12} W^2 \right)$. This stress will play the role of a residual stress. It stays negative for $\left| y \right| \leq \frac{1}{2}\frac{W}{\sqrt{3}} $ so the planar configuration is susceptible to lose stability in a central region of width $\frac{W_c}{\sqrt{3}}$, where $W_c$ is the critical width for buckling.

{\it (2) Long-range periodic configurations.} 
In rectangular coordinates, the first F{\"o}ppl-von K\'arm\'an equation predicts buckling solutions of the form\cite{ElasticityGeometryAudoly} $h(x,y) = A \cos(\beta y) \cos(q x)$, where $\beta$ fixes simultaneously the wrinkle's effective width, $D \sim \frac{1}{\beta}$, the longitudinal wavenumber, $q^4=1+\beta^4$, and the stress to order \textit{zero}, $\sigma_{xx}^{(0)}=-2 ( \beta^2 + \sqrt{1+\beta^4} )$ (see \ref{sec:toroidal2}). We will refer to those solutions simply as \textit{sinusoidal solutions}. Note that the \textit{elastica} wavenumber, $q=1$, holds only for $\beta \ll 1$ and $\sigma_{xx}^{(0)} \approx -2$. The emergent buckling domain of width $D=\frac{W_c}{\sqrt{3}}$ and stretching energy $\sim D \left[ \sigma_{xx}^{(0)} \right]^2$, should match the stretching energy of the critical planar configuration, proportional to $\int_{-D/2}^{D/2} \widetilde{\sigma}_{xx}^{\ 2} dy = \frac{\tau^2}{120} D^5$. Hence,
\begin{equation}
\frac{\tau^2}{4320} W_c^{\ 4} =  1+ 2 \beta^4 + 2 \beta^2 \sqrt{1+\beta^4} \text{.}
\label{EqForWc} 
\end{equation}

Therefore, since $\beta \sim 1/W_c$, for $\tau \ll 1$ we have $W_c \approx \frac{8.1 }{\sqrt{\tau}}$. As long as $W > W_c$, strips exhibit single  buckling domains of width $D = W - 2 w$, where $w=\frac{1}{2} ( 1 - \frac{1}{\sqrt{3}}) W_c$ is the width of the planar unbuckled zones of positive stress limiting the long edges (see the diagram in Fig.~\ref{fig:PhaseDiagram}(a) for straight wrinkles and Fig.~\ref{fig:Fig1}(c.19) for bent wrinkles). For $\tau \ll 1$, the theoretical prediction $w \approx \frac{1.70}{\sqrt{\tau}}$ agrees very well with the simulations (Fig.~\ref{fig:PhaseDiagram}(b).) Moreover, since $D \gtrsim \frac{W_c}{\sqrt{3}} \gg 1$ implies $\beta \ll 1$, the post-buckling configuration is expected to be well described by the sinusoidal solution with $q=1$. While $\tau$ is kept small, experiments and simulations corroborate strongly this theoretical framework. However, as indicated in the phase diagram in Fig.~\ref{fig:PhaseDiagram}(a), when $\tau \gtrsim 5 $, sinusoidal solutions do not appear spontaneously at the buckling transition. Instead, buckling results in periodic dimples -- where the natural topology is locally conserved -- and sinusoidal solutions emerge only for larger $W$. In practice, we classify the resulting pattern as {\it {periodic dimpling}} when the up-down symmetry is broken, so that the width of the inversions $l$ is smaller than the half-period of the pattern, $\lambda/2$ (see Fig.~\ref{fig:PhaseDiagram}(c)). Roughly speaking, dimples appear because the residual stress at the center of the strip that is proportional to $\tau W^2$, becomes critical for relatively small $W$ and $\beta \approx q$. As a result, the maximum mean curvature of the sinusoidal solution, $\frac{1}{2} A \left(q^2 + \beta^2 \right) \sim \kappa_0 W q$, is comparable to the natural curvature. Thus, a strong asymmetry (convex-concave) of the bending moment favours the local conservation of the geometry. As $W$ increases, the asymmetry becomes weaker and eventually sinusoidal profiles emerge. 

Since dimpling configurations represent a relatively narrow region of the phase space, and since long strips (longer than $\pi R$) are required to observe the intrinsic periodicity, we use simulations to characterise them systematically, over a wide range of $\tau$. In particular, we find that (for $\tau \gtrsim 5 $) sinusoidal solutions appear after a continuous transition terminating approximately at $W \approx 2\pi$.

A rigorous study of the dimpling mechanics is out of the scope of this paper. However, we can estimate the buckling threshold $W_c(\tau)$ by assuming that the stresses of the dimples are comparable to those of other profiles (to order zero). Taking $h(x , \pm D/2)=0$ sets $\beta = \frac{\pi \sqrt{3}}{W_c}$ and from Eq.~\ref{EqForWc} we obtain the numerical solution $W_c(\tau)$, plotted in Fig.~\ref{fig:PhaseDiagram}(a) with a green line. This solution gives a good global estimate for the buckling transition.  

Due to (and despite) the intermediate dimpling regime the sinusoidal solutions satisfy $\beta \ll q$ even when $\tau \gg 1$. Thus, wrinkling configurations are always characterised by $q = 1$ and $\sigma_{xx}^{(0)} = -2$. Up to order $\kappa_0^{\ 2}$ and using the residual stress decomposition, $T_{ik}  \rightarrow \sigma_{ik} - \widetilde{\sigma}_{ik}$, the second order approximation of the principal stresses is (\ref{sec:toroidal2}), 
\begin{equation}
\sigma_{xx} = -2 + \tau O(y^4) \text{, \ } \sigma_{yy} = \frac{\tau}{4} \cos(2x) \text{.}
\label{AnisotropicStressSolution} 
\end{equation} 
Those solutions are shown in Fig.~\ref{fig:Fig1}(b.2), along with the stresses measured in finite element simulations. As in the axisymmetric configurations, the stress oscillates with period $\lambda/2$ and an amplitude proportional to $\tau$. However, there are important differences. The transverse stress of straight wrinkles ($\sigma_{yy}$) oscillates around zero, independently of the wrinkle amplitude, while for axisymmetric wrinkles the hoop stress oscillates around $-2$ with an intensity coupled to the local amplitude $A \sim r$ (see Eq.\ref{StressResultAxissymm}). As samples get bigger, the increasingly large stresses of axisymmetric wrinkles lead to a loss of stability, favouring straight wrinkles.

{\it (3) Domain reconfiguration and folds.}
Similarly to planar films subjected to compressive stresses \cite{Pocivavsek2008, Kim2011, Oshri2015}, floating shells exhibit a wrinkle to fold transition when the confinement or excess length increases sufficiently. The strain localizes into folds whose amplitude grows until they make self-contact \cite{Holmes2010, Kim2011, Brau2013}.

In contrast to elasticas and planar sheets on liquid substrates, folding is not accompanied by flattening of the remainder of the sample. Rather, we observe domain reconfigurations, such that folds become boundaries of newly formed wrinkling domains with amplitude $A=\frac{\kappa_0 D}{q \sqrt{2}}$. This relationship is valid up to a critical maximum width $D \approx 6.0/\kappa_0$ (see Fig.\ref{fig:PhaseDiagram}(d) where $\tau$ has been chosen large enough so that $w$ is negligibly small, and we can take $D=W$) setting the critical amplitude for folding around $A \approx 0.5$. Typically, prior to a fold transition, we observe breaking of the long-range order of the wrinkling pattern within the domains, as shown in Figs.~\ref{fig:PhaseDiagram}(a) and~\ref{fig:WTF-slices}. This transformation is characterised by the growth of the amplitude within the main domain, as demonstrated in Fig.~\ref{fig:PhaseDiagram}(d) for series of experiments and simulations of samples with high aspect ratio. An analogous transition occurs in samples with small aspect ratio, \textit{e.g.} in caps, hexagons and squares (see Fig.~\ref{polygons}).


\section*{Discussion}

We have proposed a way of generating controllable and reproducible patterns in thin elastic shells, with potential applications in technology and engineering. For example, by using UV-curable adhesives (\textit{e.g.} NOA, \cite{Bartolo2012}), one could take advantage of the discussed mechanisms of pattern formation to produce microfluidic stickers for complex microfluidic devices, patterned substrates for directed cell growth experiments, or pressure sensors \cite{Genzer2006, jiang2002controlling}. The intrinsic curvature stabilizes the pattern, allowing for it to be maintained without the continuous application of external forces. Moreover, using this technique we can precisely control the spatial variation of the amplitude, both globally and locally.  If the shell's properties can be changed dynamically (\textit{e.g.} by heat, light, or electromagnetic fields), then additional applications of adaptable surfaces are possible, for example with microscale features creating responsive friction or adhesion.  

\begin{figure}[b]
\centering
\includegraphics[width=1.0\linewidth]{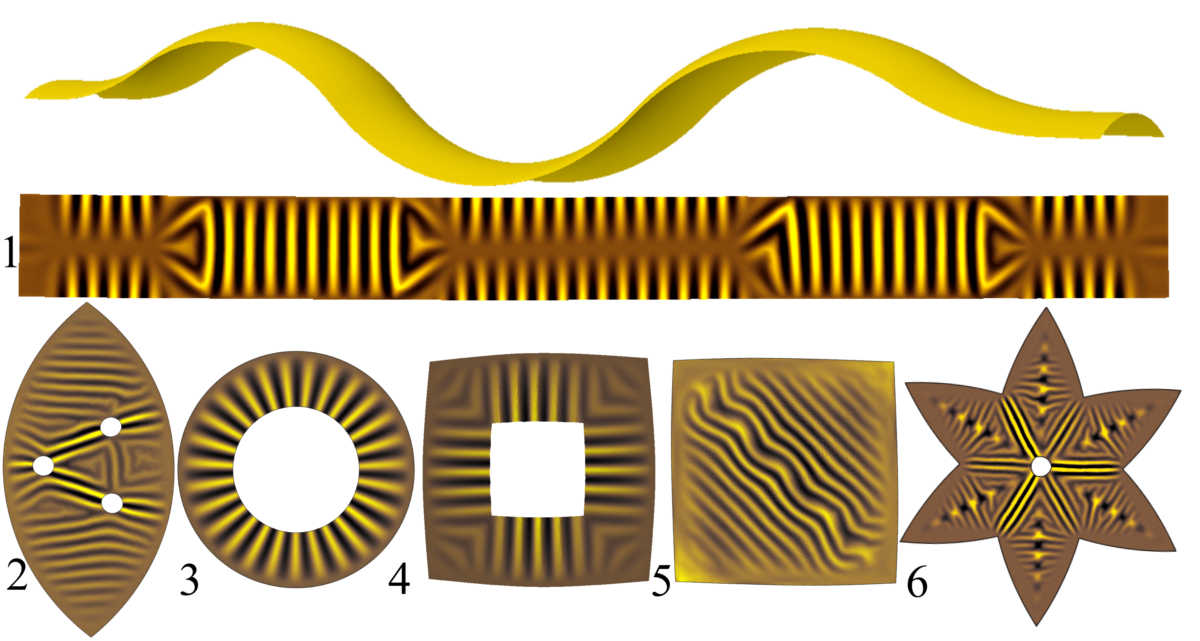}
\caption{Variations in curvature, thickness and material properties allow for precise control over wrinkle structures.  Shown are: (1) a strip with longitudinal curvature variation, as shown in the undeformed configuration at the top; (2) holes interacting in a spherical slice; (3) a spherical annulus; (4) a square frame; (5) a stiff ray along the diagonal of a square shell resulting in a wavy pattern, in contrast to the homogeneous case shown in Fig.~\ref{fig:Fig1}(c.2,c14); (6) a spherical ``flower'' with stiff rays connecting the hole and the tips of the ``petals''. The colouring corresponds to the vertical coordinate.}
\label{fig:Designs}
\end{figure}

Introducing additional geometrical and compositional complexity into this system would also open the door to a strikingly rich phenomenology of patterns, allowing for further self-assembling structures.  Using curvature variation and purposely introduced inhomogeneities, such as local variations of the elastic properties and thickness, one can design in a bespoke amplitude, direction, domain size and orientation of a wrinkle pattern. For this, a combination of the instabilities discussed in this work (wrinkles, dimples, unbuckled regions and folds) can be harnessed to achieve intricate patterns. Several examples are shown in Fig.~\ref{fig:Designs}, but the space of patterns is immense.  

So far, we have focussed on examples where the intrinsic curvature is positive and constant, \textit{i.e.}, spherical shells. Naturally, one could also consider scenarios where the Gaussian curvature is negative (hyperbolic shells), or is varied across the sample, and the mathematical techniques that we have explored can be applied to these situations as well. One such example is shown in Fig.~\ref{fig:Designs}(1) and has alternating regions of positive and negative Gaussian curvature. As a floating shell it has wrinkles of constant amplitude, arranged in domains. For negative curvature, we see a flat region along the sample's longitudinal midline, while for positive curvature, the flat regions are along the edges. This effect can be explained with the concept of residual stress (see Eq.~\ref{ResidualStrainX2}), which in a toroidal strip of Gaussian curvature $\kappa_G$ takes the form $\widetilde{\sigma}_{xx} = \frac{3 \eta }{8} \kappa_G \left( y^2 - \frac{1}{12} W^2 \right) $. So, depending on the sign of $\kappa_G$, compression ($\widetilde{\sigma}_{xx} < 0$) may appear at a central region or close to the edges. Since the local compression drives wrinkling, the sign of the natural Gaussian curvature determines the position of the unbuckled zones. 

Finally, material inhomogeneities and holes have been shown to allow for precise control of the wrinkle orientation in flat double-layered systems \cite{bowden1998spontaneous, hendricks2007wrinkle, Wu2014}. The coupling of such inhomogeneities with the curvature-induced confinement in floating shells can produce striking patterns; some examples with holes and/or stiffness contrasts are shown in Fig.~\ref{fig:Designs}(2-6).  The precise mechanisms of shape control in these more complex shells remain to be explored, but can be expected to further establish the floating shell as an innovative tool in the arsenal of engineered methodologies for pattern formation.

\subsection*{METHODS}
\small{

\subsection*{Preparation of PDMS shells}
The shells were fabricated using hemispherical acrylic moulds from 4 to 15 cm in radius. A curable elastomer (PDMS, Sylgard 184, Dow Corning) was freshly prepared and poured into the moulds to coat their inner surface. The coated moulds were then suspended upside-down for at least 24 hours, so that the liquid PDMS drained under the effect of gravity. This viscous drainage, accompanied by the slow room-temperature curing of the polymer, created a solid elastic shell of relatively even thickness \cite{Lee2016} that could be peeled away from the mould. The minimum shell thickness produced was about 30 $\mu$m. Thicker layers were made by repeated application of PDMS, followed by drainage after each layer. 

Shells were nominally made with a 10:1 ratio of polymer base to cross-linker, resulting in a Young's modulus of $E$ = 1.5 MPa.  Softer shells, down to $E$ = 20 kPa, were obtained by decreasing the cross-linker concentration.  The mechanical properties of bulk PDMS samples, prepared and cured under identical conditions, were measured by uniaxial compression tests and reported in Ref.~\cite{Hemmerle2016}.  

To facilitate the manipulation of the softest shells ($e.g.$ 30-60 $\mu$m and low $E$), we pre-treated the moulds by adding a coating of water-soluble PVA glue, deposited by the same techniques as described above, and dried for 60 minutes.  The resulting glue-PDMS bilayers were easier to handle and cut, and the glue dissolved after the shells were transferred onto a water substrate.  The last traces of wet glue were removed by mechanical agitation, for example with a small brush, or by gently rubbing against a container wall.

To evolve a wrinkled state, a shell was peeled from its mould, cut into the desired shape (\textit{e.g.} a spherical strip of constant width), and floated on a bath of water.  Profiles of the deformed shell were made by projecting a vertical laser sheet across the shell, and taking calibrated digital side-view images of the scattered laser light (Nikon D5100 digital SLR camera).  Additionally, the full wrinkle pattern was imaged by shadowgraph techniques: collimated light was projected through the floating shell from below, onto a translucent screen positioned just above the shell.  The wrinkle troughs and ridges acted to focus or diverge light, resolving a pattern of bright and dark stripes on the screen, which was then imaged digitally. Note that the resulting colour scale is not linear, hence the  starker appearance of the regions with high positive values of the vertical deflection when compared to the simulations results, where a linear colour scale was used.


\subsection*{Finite Element Modelling}
Numerical simulations were performed using the commercial finite element package ABAQUS/Explicit.
Four-node thin shell elements with reduced integration (element type S4R) were used in all simulations and a mesh sensitivity study verified that the results were independent of the element size, as long as the element size is at least one order smaller than the wavelength of the final pattern. The gravitational load was specified as a non-uniform distributed load over the shell surface, using a VDLOAD subroutine. Free boundary conditions were used in all simulations.
We performed comparative non-linear geometric Finite Element Analysis (FEA) using both linear elastic and Neo-Hookean hyper-elastic materials and the results were not dependent on the choice of elastic model. 
The colour-coding on all of the FEA images corresponds to the vertical deflection of the shell from the (initially flat) surface of the liquid.


\section*{Acknowledgements}
O.A. received support from La Comisi\'on Nacional de Investigaci\'on Científica y Tecnol\'ogica (CONICYT) through the program Becas de Postdoctorado en el Extranjero BECAS CHILE – Convocatoria 2017. D.V.T. and E.K. were supported by the NSF Award PHY-1554887, NSF MRSEC Grant DMR-1120901 and the Burroughs Wellcome Career Award.

\clearpage
\clearpage
\renewcommand{\thepage}{S\arabic{page}} 
\renewcommand{\thesection}{S\arabic{section}}  
\renewcommand{\theequation}{S\arabic{equation}}  
\renewcommand{\thefigure}{S\arabic{figure}} 
\setcounter{page}{0}    
\setcounter{section}{0}    
\setcounter{figure}{0}    
\setcounter{equation}{0}    

\section{Supplementary Results}

\subsection{Axisymmetric Energy Functional}

The local state of deformation (at a given arc-length position $s$) of the shell will be defined through the radial coordinate, $r$, the vertical position, $h$, and the tangential angle, $\theta$ (see Fig.\ref{Parametrization}). We use the subscript \textit{zero} to denote the non-deformed state reference of the shell and we parametrize the system with the natural meridional position $s_0$. 

\begin{figure}[htbp]
\centerline{\includegraphics[scale=0.19]{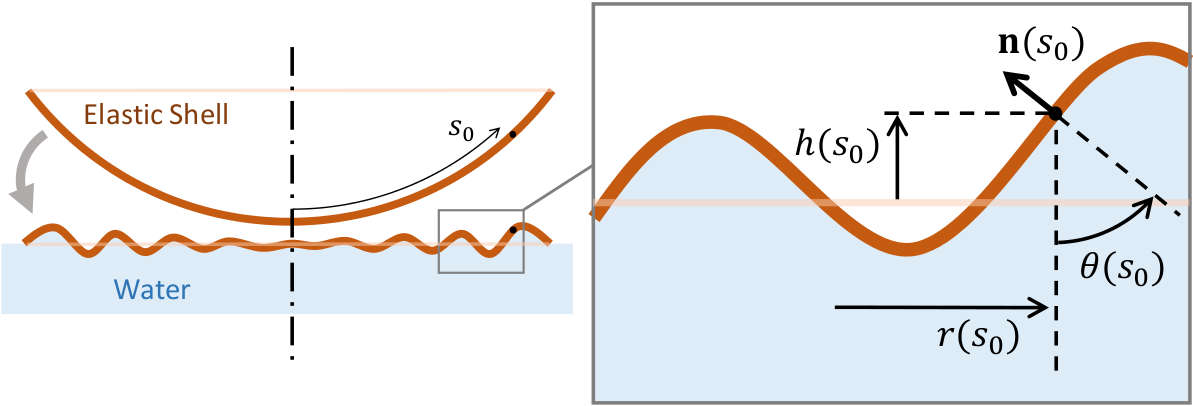}}
\caption{Parametrization of the floating shell in the axisymmetric configuration.}
\label{Parametrization}
\end{figure}

The surface density of pure bending energy of a plate without prescribed curvature is
\begin{equation}
\mathcal{F}_B= \frac{1}{2} K_B \left( \kappa_1 + \kappa_2 \right)^{2} + \frac{1}{2} K_G  \left( \kappa_1 \kappa_2 \right)  
\label{BendingEnergyPlate} \text{,}
\end{equation}
where $\kappa_1$, $\kappa_2$, are the principal curvatures, $K_B$ is the bending stiffness and $K_G$ is the Gaussian rigidity. For conventional materials with Poisson's ratio $1/2$, the plate theory predicts $K_B = -K_G =\frac{1}{9} Et^{3}$, where $E$ is the Young's modulus and $t$ is the thickness. For a surface of revolution the principal curvatures are spatial derivatives of the normal vector, $\textbf{n}$, in the direction of the side profiles (meridians) and the hoop direction (parallels). Thus, the principal curvatures are $\kappa_1=\frac{d \theta}{d s} $ and $\kappa_2=\frac{\sin \theta}{r}$. Similarly, the principal natural curvatures are $\frac{d \theta_0}{d s_0}$ and $\frac{\sin \theta_0}{r_0}$. When considering axisymmetric shells, the principal curvatures $\kappa_1$ and $\kappa_2$ in Eq.~\ref{BendingEnergyPlate} need to be replaced by the curvature difference from the undeformed state $\Delta\kappa_1 = \kappa_1-\frac{d \theta_0}{d s_0}$ and $\Delta\kappa_2 = \kappa_2-\frac{\sin \theta_0}{r_0}$ respectively. Then, $\kappa_1 = \frac{1}{1+\epsilon_1} \theta' - \theta_0'$ and $\kappa_2 = \left( \frac{\sin \theta}{1+\epsilon_2} - \sin \theta_0 \right) \frac{1}{r_0} $, where the apostrophes denote derivatives respect to $s_0$, $\epsilon_1 = \frac{ds-ds_0}{ds_0}=\sqrt{r'^2 + h'^2}-1$, is the meridional strain and $\epsilon_2 = \frac{r-r_0}{r_0}$ is the hoop strain.

We assume the shell is shallow and the deflections are small, so $\theta_0 \ll 1$ and $h' \ll 1$. In general, $h'= \left( 1+ \epsilon_1 \right) \sin \theta$. Further, $\sin \theta \approx \theta$ and at leading order $\sin \theta = h'$ and $\theta' = h''$.  Since the vertical displacements are dominant compared to the in-plane deformations, up to second order we have $\Delta\kappa_1 =  h'' - \theta_0'$ and $\Delta\kappa_2= \frac{h' - \theta_0}{r_0}$. Then, up to constant terms we have 
\begin{equation}
\begin{split}
\mathcal{F}_B= & \frac{1}{2} K_B \left[ h''^2 - 2 \theta_0' h'' + \frac{h'^2 - 2 \theta_0 h'}{ r_0^{\ 2}}  \right]  + \\ & + \frac{1}{4} K_B \frac{\left( h'^2 - 2 \theta_0 h' \right)'}{r_0} \text{.}
\end{split}
\label{BendingEnergy} 
\end{equation} 
The total bending energy is computed by the surface integral $\int \mathcal{F}_B \ r_0 \ ds_0$. Therefore, the last term of Eq.\ref{BendingEnergy} can be  dropped, as well, because its associated integral is a boundary value.
 
Similarly to Eq.\ref{BendingEnergyPlate}, the in-plane stretching energy density $\mathcal{F}_S$ for linear and isotropic materials reads
\begin{equation}
\mathcal{F}_S= \frac{1}{2} Y_S \left(\epsilon_1 + \epsilon_2 \right)^{2} + \frac{1}{2} Y_G  \left( \epsilon_1 \epsilon_2 \right)  
\label{StretchingEnergy} \text{,}
\end{equation}   
where $Y_S$, $Y_G$, are the elastic rigidities. Since the Poisson's ratio equals $1/2$ we take $Y_S=-Y_G=\frac{4}{3}Et$. 

The hydrostatic energy per unit of non-deformed surface of the shell $\mathcal{F}_H$, represents the mechanical work done by the hydrostatic force $-\left(1+\epsilon_2 \right) r' K_g h$, where $K_g$ is the specific weight of the liquid. Thus, the density of hydrostatic energy, for small deflections and infinitesimal strain, is
\begin{equation}
\mathcal{F}_H = \frac{1}{2} \ K_g h^2 
\label{TotalHydrostaticEnergy} \text{.}
\end{equation}

The total energy density $\mathcal{F} = \mathcal{F}_{B} + \mathcal{F}_{S} + \mathcal{F}_{H}$ depends on the mechanical stiffnesses $K_B \sim E t^3$, $K_g$, $Y_S \sim Et$, and the natural geometry of the shell, $\theta_0(s_0)$. We normalize all lengths by $\Lambda \equiv \left( K_B / K_g\right)^{1/4}$ and all energy densities by $ K_g \Lambda^2$. Then, in dimensionless form, we have
\begin{equation}
\begin{split}
\mathcal{F} & =  \frac{1}{2} h''^2 - \theta_0' h'' + \frac{h'^2 - 2 \theta_0 h'}{2 r_0^{\ 2}} +  \frac{1}{2} h^2 + \\ & + \frac{\eta}{2} \left( \epsilon_1^{\, 2} + \epsilon_2^{\, 2} + \epsilon_1 \epsilon_2 \right)
\label{DimensionLessTotalEnergy2} \text{,}
\end{split}
\end{equation} 
where $\eta \equiv \frac{ Y_S}{K_g \Lambda^2} \sim \left( \Lambda / t \right)^2 \ggg 1$ is the normalized stretching rigidity. 

The coupling between the local deflections and the in-plane deformation can be obtained directly from the definitions of the strain elements ($\epsilon_1$, $\epsilon_2$). Considering $r'^2 + h'^2 = \left( 1+ \epsilon_1\right)^2  $, up to second order deflections we get 
\begin{equation}
r'=1+\epsilon_1 - \frac{1}{2} h'^2
\label{rPrime} \text{.}
\end{equation} 
The derivative of the hoop strain is $r'=r_0' + \left( r_0 \epsilon_2 \right)'$ and from Eq.\ref{rPrime} we get the deflection-stretching relationship,
\begin{equation}
\frac{1}{2} h'^2 - \epsilon_1 + \left( r_0 \epsilon_2 \right)' = 1- r_0'    
\label{DeflectionStretching} \text{.}
\end{equation}  
Note that $1- r_0'$ is the dimensionless local lateral excess of length imposed by the shell, denoted by $\varepsilon$ in the main text. 

Since the in-plane displacements are infinitesimal, the coordinates $(r, s)$ and $(r_0, s_0)$ are interchangeable to the lowest order. Then, $h'=r_0' \frac{d h}{dr}$ and the higher order derivatives are
\begin{equation}
\begin{split}
h''= & r_0'' \frac{d h}{dr} + r_0'^2 \frac{d^2 h}{dr^2}  \text{\ ,} \\
h'''= & r_0''' \frac{d h}{dr} + 3 r_0' r_0'' \frac{d^2 h}{dr^2} + r_0'^3 \frac{d^3 h}{dr^3}  \text{\ ,} \\
h''''= & \left( 4 r_0' r_0''' + 3r_0''^2 \right) \frac{d^2 h}{dr^2} + 6r_0'^2 r_0'' \frac{d^3 h}{dr^3} + r_0'^4 \frac{d^4 h}{dr^4} \text{\ .}
\label{h''_h'''_h''''}
\end{split}
\end{equation}
Calculating the derivatives of $r_0$ and $\theta_0$ for Eq.~\ref{h''_h'''_h''''},   we have taken into account that the longitudinal natural curvature $\theta_0'=\kappa_0 \ll 1$ changes very slowly. Since for shallow shells we have $r_0' = \sqrt{1-h_0'^2} \approx 1 - \frac{1}{2} \theta_0^{\ 2}$, the second and third derivatives of $r_0$ are 
\begin{equation}
r_0''= - \kappa_0 \theta_0 \text{\ , \ \ \ } r_0''' = - \kappa_0^{\ 2}
\label{r0_r0''_r_0'''} \text{,}
\end{equation}
while the forth derivative, $r_0''''$ have been set to 0. 
In terms of $\kappa_0$ and $\theta_0$, the $h$ derivatives are
\begin{equation}
\begin{split}
h'' \approx  &  \ddot{h} -  \kappa_0 \theta_0 \dot{h} \text{\ , \ } h''' \approx  \dddot{h} - 3 \kappa_0 \theta_0 \ddot{h} \text{\ ,} \\
h'''' \approx  &  \ddddot{h} -  4 \kappa_0^{\ 2} \ddot{h} - 6  \kappa_0 \theta_0 \dddot{h}  \text{\ ,}
\label{h''_h'''_h''''_b}
\end{split}
\end{equation}  
where the over-dots denote $r$-derivatives. 

In the regime studied in this work the wrinkling patterns have local amplitude proportional to the square-root of the excess of length $\varepsilon = 1- r_0' \approx \frac{1}{2} \theta_0^{\ 2}$. Thus, we can assume $h \sim \theta_0$, and up to order $\theta_0^{\ 2}$ the total energy (Eq.\ref{DimensionLessTotalEnergy2}) is
\begin{equation}
\begin{split}
\mathcal{F} & = \frac{1}{2} \left[  \ddot{h}^2 + \left( \frac{\dot{h}}{r} \right)^2 - 2 \left( \kappa_0 \ddot{h} + \frac{\theta_0 \dot{h}}{r^2} \right) \right]  +  \frac{1}{2} h^2 + \\ & + \frac{\eta}{2} \left( \epsilon_1^{\, 2} + \epsilon_2^{\, 2} + \epsilon_1 \epsilon_2 \right)  \text{.}
\end{split}
\label{DimensionLessTotalEnergy3} 
\end{equation}   
Moreover, up to order $\theta_0^{\ 2}$, Eq.\ref*{DeflectionStretching} becomes
\begin{equation}
\varepsilon= \frac{1}{2} \theta_0^{\ 2} = \epsilon_2 + r \dot{\epsilon}_2 - \epsilon_1 + \frac{1}{2} \dot{h}^2     
\label{DeflectionStretching2} \text{.}
\end{equation} 

Further, we assume that the natural curvature $\theta_0'=\kappa_0(r)$ is a smooth function. For spherical shells $\kappa_0$ is constant and since $\theta_0' = \left[  1 - \frac{1}{2} \theta_0^{\ 2} + O(\theta_0^{\ 4}) \right]  \frac{d}{dr} \theta_0$, we have $\theta_0 = \kappa_0 r + \frac{1}{6} \theta_0^{\ 3} + O(\theta_0^{\ 5})$ and, up to second order, $ \theta_0 = \kappa_0 r$. Substituting in Eq.\ref{DimensionLessTotalEnergy3} and Eq.\ref{DeflectionStretching2}, we obtain the relationships for $\mathcal{F}$ and $\varepsilon$ used in the main text.  

\subsection{Minimization of the Energy Functional}

After the transformation $ \mathcal{F} \rightarrow  \mathcal{F} + \mu(r) \varepsilon $, where $\mu(r)$ is a local Lagrange multiplier ensuring the constraint $\varepsilon= \frac{1}{2} \theta_0^{\ 2} $ (Eq.\ref{DeflectionStretching2}), the energy functional writes, 
\begin{equation}
U = \int_{r_i}^{r_f} \mathcal{F}(r, \epsilon_1, \epsilon_2, \dot{\epsilon}_2,  h,\dot{h},\ddot{h}) \, r \, dr 
\label{Functional} \text{.}
\end{equation}
The problem of minimization is reduced to solving a system of three differential equations.

Since the density energy we consider here is a particular case of $\mathcal{F}=\mathcal{F}(r, \epsilon_1, \dot{\epsilon}_1, \ddot{\epsilon}_1, \epsilon_2, \dot{\epsilon}_2, \ddot{\epsilon}_2, h, \dot{h}, \ddot{h})$, the necessary condition for minimization can be generalized from the study of a functional of the form $G= \int_{r_i}^{r_f} \mathcal{L}(r, \text{g}, \dot{\text{g}}, \ddot{\text{g}} ) \, r \, dr $, where $\text{g}(r)$ represents any of the independent functions ($\epsilon_1$, $\epsilon_2$, $h$). The necessary condition for minimisation is 
\begin{equation}
\begin{split}
\delta G &= \int_{r_i}^{r_f} r \, \delta \mathcal{L} \, dr = 0 \text{ \ with \ } \\ \delta \mathcal{L} = & \mathcal{V} \ \delta \text{g}  + \frac{d}{dr}\left[ \mathcal{W} \ \delta \text{g} + \mathcal{Z} \ \delta \dot{\text{g}} \right]  \text{,}
\label{Functional2} 
\end{split}
\end{equation}
where $\mathcal{V}=\frac{\partial \mathcal{L}}{\partial \text{g}} - \frac{d}{d r}\left(\frac{\partial \mathcal{L}}{\partial \dot{\text{g}}} \right) +  \frac{d^2}{dr^2}\left(\frac{\partial \mathcal{L}}{\partial \ddot{\text{g}}} \right)$, $\mathcal{W}= \frac{\partial \mathcal{L}}{\partial \dot{\text{g}}} - \frac{d}{dr}\left(\frac{\partial \mathcal{L}}{\partial \ddot{\text{g}}} \right)$ and $\mathcal{Z}=\frac{\partial \mathcal{L}}{\partial \ddot{\text{g}}}$. Thus, 
\[
r \, \delta \mathcal{L} =  \left[ r \mathcal{V} - \mathcal{W} +  \dot{\mathcal{Z}} \right]  \delta \text{g} + \frac{d}{dr}\left[ \left( r \mathcal{W} - \mathcal{Z} \right) \delta \text{g} + r \mathcal{Z} \delta \dot{\text{g}} \right] \text{.}
\]
Then, assuming fixed ends, the functional derivative of the energy vanishes only when the condition $r \mathcal{V} - \mathcal{W} +  \dot{\mathcal{Z}} = 0 $ is satisfied. Therefore,
\begin{equation}
\begin{split}
r & \left\lbrace \frac{\partial \mathcal{L} }{\partial \text{g}} - \frac{d}{dr}\left( \frac{\partial \mathcal{L} }{\partial \dot{\text{g}}}\right) + \frac{d^2}{dr^2}\left( \frac{\partial \mathcal{L} }{\partial \ddot{\text{g}}}\right) \right\rbrace  + \\  & + 2 \frac{d}{dr}\left( \frac{\partial \mathcal{L} }{\partial \ddot{\text{g}} }\right) - \frac{\partial \mathcal{L} }{\partial \dot{\text{g}}}  = 0  \text{.}
\end{split}
\label{ELEquation} 
\end{equation}

For $\text{g}=\epsilon_1$ and $\mathcal{L} = \mathcal{F}(r, \epsilon_1, \epsilon_2, \dot{\epsilon}_2,  h,\dot{h},\ddot{h})$, Eq.\ref{ELEquation} writes $\frac{\partial \mathcal{F}}{\partial \epsilon_1} = 0$. Then $\mu = \eta \left(  \epsilon_1 + \frac{1}{2} \epsilon_2 \right)$, so the Lagrange multiplier equals the meridional stress, $\sigma_1= \mu$. For $\text{g}=\epsilon_2$ we find $ \frac{\partial \mathcal{F}}{\partial \epsilon_2} - \frac{d}{dr} \left( \frac{\partial \mathcal{F}}{\partial \dot{\epsilon}_2 }\right) = \frac{1}{r} \frac{\partial \mathcal{F}}{\partial \dot{\epsilon}_2 }  $. Therefore,
\begin{equation}
\sigma_2 = \sigma_1 + \frac{d}{dr} \left( r \sigma_1 \right) 
\label{StressBalance} \text{,}
\end{equation}
where $\sigma_2=\eta \left(  \epsilon_2 + \frac{1}{2} \epsilon_1 \right)$ is the hoop stress. Substituting $\epsilon_1= \frac{4}{3 \eta} \left( \sigma_1 - \frac{1}{2} \sigma_2 \right) $, $\epsilon_2= \frac{4}{3 \eta} \left( \sigma_2 - \frac{1}{2} \sigma_1 \right) $ in Eq.\ref{StressBalance} yields
\begin{equation}
r^2 \ddot{\sigma}_1 + 3r \dot{\sigma}_1 = \frac{3 \eta }{8} \left( 2 \varepsilon - \dot{h}^2 \right) 
\label{StressBalance2} \text{.}
\end{equation}
This equation is identical to Eq.\ref{StressEquation} in the mean text. Finally, for $\text{g}=h$, Eq.\ref{ELEquation} becomes
\begin{equation}
\begin{split}
r & \left\lbrace \frac{\partial \mathcal{F} }{\partial h} - \frac{d}{dr}\left( \frac{\partial \mathcal{F} }{\partial \dot{h}}\right) + \frac{d^2}{dr^2}\left( \frac{\partial \mathcal{F} }{\partial \ddot{h}}\right) \right\rbrace  + \\  & + 2 \frac{d}{dr}\left( \frac{\partial \mathcal{F} }{\partial \ddot{h} }\right) - \frac{\partial \mathcal{F} }{\partial \dot{h}}  = 0
\end{split}
\label{hELEquation} 
\end{equation}
and then 
\begin{equation}
\ddddot{h} - \sigma_1 \ddot{h} + h + \frac{2\dddot{h} - \sigma_1 \dot{h}}{r} =  \frac{\ddot{h}}{r^2}  + \left( \dot{\sigma_1} - \frac{1}{r^3} \right) \dot{h}  
\label{DeflectionEquation2} \text{\ ,}
\end{equation}
which is exactly Eq.\ref{DeflectionEquation}.

\subsection{Axisymmetric F{\"o}ppl-von K\'arm\'an equation} 

In polar coordinates the first F{\"o}ppl-von K\'arm\'an equation is \cite{King2012}
\begin{equation}
\begin{split}
K_B & \Delta^2 h  - \sigma_{rr} \partial_r^2 h - \frac{2}{r} \sigma_{r\varphi} \left( \partial_r - \frac{1}{r} \right) \partial_{\varphi} h + \\ 
- & \frac{1}{r^2} \sigma_{\varphi \varphi}\left( \partial_{\varphi}^2 h + r \partial_r h \right) = F_N \text{\ ,}
\end{split} 
\label{vonKarmanEquations} 
\end{equation} 
where $\varphi$ is the polar angle, $\Delta= \partial_r^2 + \frac{1}{r} \partial_r + \frac{1}{r^2} \partial_{\varphi}^2$ is the Laplacian, $\sigma_{ij}$ are in-plane stress elements and $F_N=- K_g h$ is the normal external force per unit surface exerted by the liquid. We use $\Lambda = \left( K_B / K_g \right)^{1/4} $ as a normalization length scale, and $K_g \Lambda^2$ to nondimensionalise the stresses. Also, for axisymmetric configurations $\Delta^2 =\partial_r^4 + \frac{2}{r} \partial_r^3 - \frac{1}{r^2} \partial_r^2 + \frac{1}{r^3} \partial_r$, then for $\sigma_{rr}=\sigma_1$ and $\sigma_{\varphi \varphi}=\sigma_2$, Eq.\ref{vonKarmanEquations} leads to Eq.\ref{DeflectionEquation2}.

\subsection{F{\"o}ppl-von K\'arm\'an equations in rectangular coordinates} 

The dimensionless form of the first F{\"o}ppl-von K\'arm\'an equation in rectangular coordinates is 
\begin{equation}
\Delta^2 h  + \left( 1 - \sigma_{xx} \partial_x^2 - 2 \sigma_{xy} \partial_{xy}^2 - \sigma_{yy} \partial_y^2 \right) h   =0 \text{,} 
\label{vonKarmanEquationsRectangular} 
\end{equation} 
where $\Delta = \partial_x^2 + \partial_y^2$ is the Laplacian. The stress components  have to satisfy the equilibrium equations
\begin{equation}
\partial_x \sigma_{xx} + \partial_y \sigma_{xy} = 0 \text{ \  , \ } \partial_x \sigma_{xy} + \partial_y \sigma_{yy} = 0 \text{.}
\label{StressEquilibriumRectangular} 
\end{equation}  
In terms of the strain components,    
\begin{equation}
\begin{split}
\sigma_{xx}= \eta \left( \epsilon_{xx} + \frac{1}{2} \epsilon_{yy}\right)  \text{,\ } & \sigma_{yy} = \eta \left( \epsilon_{yy} + \frac{1}{2} \epsilon_{xx}\right) \text{,} \\ \sigma_{xy} &= \dfrac{\eta}{2} \epsilon_{xy} \text{.}
\end{split}
\label{StressComponentsRectangular} 
\end{equation} 
Similar to the axisymmetric case, the deflection-stretching relationship can be inferred from the definition of the strains. Denoting the components of the displacement field $\text{u}_x$, $\text{u}_y$, $\text{u}_z =h(x,y)$, the strain elements can be written as 
\begin{equation}
\begin{split}
\epsilon_{xx} = & \partial_x \text{u}_x + \frac{1}{2} \left( \partial_x h \right)^2    \text{ \  , \ } \epsilon_{yy} = \partial_y \text{u}_y + \frac{1}{2} \left( \partial_y h \right)^2  \text{ ,} \\
&\epsilon_{xy} = \frac{1}{2} \left( \partial_y \text{u}_x + \partial_x \text{u}_y \right) + \frac{1}{2} \left( \partial_x h \right) \left( \partial_y h \right)  \text{,}      
\end{split}
\label{StrainComponentsRectangular} 
\end{equation} 
where we have kept only non-linear terms proportional to the out-of-plane displacements, $h(x,y)$. 

Using the Airy potential, $\chi(x,y)$, one can combine Eqs.\ref{StressComponentsRectangular} and Eqs.\ref{StrainComponentsRectangular} into a single expression for the stresses and the deflections. The Airy potential, by definition, contains general solutions of Eqs.\ref{StressEquilibriumRectangular},
\begin{equation}
\sigma_{xx}=\partial_y^{2} \chi \text{ \ , \ \ } \sigma_{xy}=-\partial_{xy}^{2} \chi \text{ \ , \ \ } \sigma_{yy}=\partial_x^{2} \chi \text{.}
\label{AiryPotential} 
\end{equation} 
Therefore, the stress equilibrium can be reduced to
\begin{equation}
\Delta^2 \chi = \eta \left[ \left( \partial_{xy}^2 h \right)^2 - \left( \partial_x^2 h \right) \left( \partial_y^2 h \right)  \right] \text{.}
\label{StressBalanceAiryPotential} 
\end{equation} 

\subsection{Unbuckled configurations of toroidal strips} \label{sec:toroidal1}

We define a \textit{spherical} strip as a long rectangular sector of width, $W$, cut out of a spherical shell with curvature $\kappa_0$, aligned along the equator (see Fig.\ref{sketch2}). The mid-line of the strip is parallel to the $x$-axis so it remains at $y=0$ and the long edges -- at $y=\pm W/2$. The cross sections of the \textit{spherical} strips define meridional lines of  curvature $\kappa_1$, equal to the curvature $\kappa_2$ of the central line ($\kappa_1=\kappa_2=\kappa_0$). As a generalization of the spherical geometry, we will consider \textit{toroidal} strips where $\kappa_1 \neq \kappa_2$.  

\begin{figure}[htbp]
\centerline{\includegraphics[scale=0.8]{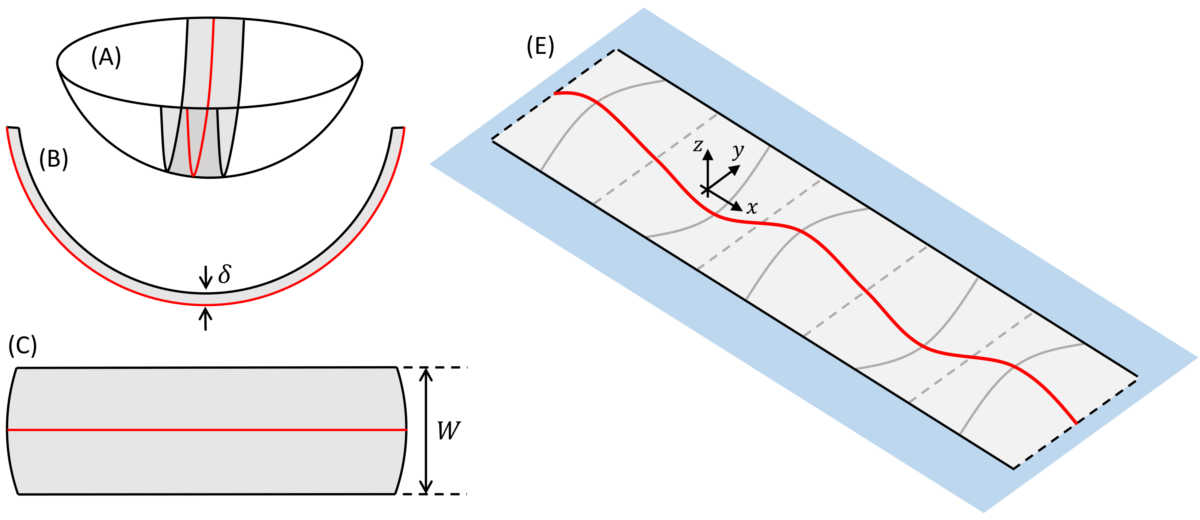}}
\caption{Representation (A-C) of the undeformed state and the respective floating configuration (E) of a spherical strip of width $W$, projected thickness $\delta$ and radius of curvature $R$. (A), (B), (C), show respectively a sketch of the isometric projection of the strip, the side view of the strip and top view of the strips.}
\label{sketch2}
\end{figure} 

The planar configuration of the strip can be determined as follows. When the strip is bent and kept straight along the $x$-axis, similarly to deflection of beams, the longitudinal strain takes the form $\widetilde{\epsilon}_{xx} = \kappa_2 z(y) - \epsilon$, where $z(y) \approx \frac{1}{2} \kappa_1 y^2$ is the natural cross-section profile and $\epsilon$ is the longitudinal strain at the centre of the strip. The resultant configuration has vanishing Gaussian curvature. Assuming that meridional and hoop lines are principal directions with stresses $\widetilde{\sigma}_{xx}=\eta \left( \widetilde{\epsilon}_{xx} + \frac{1}{2} \widetilde{\epsilon}_{yy} \right)$, $\widetilde{\sigma}_{yy}=\eta \left( \widetilde{\epsilon}_{yy} + \frac{1}{2} \widetilde{\epsilon}_{xx} \right)$, the density of the stretching energy is $\mathcal{F}_S= \frac{\eta}{2} \left(\widetilde{\epsilon}_{xx}^{\ 2} + \widetilde{\epsilon}_{xx} \widetilde{\epsilon}_{yy} + \widetilde{\epsilon}_{yy}^{\ 2} \right)$. When the configuration is such that an arbitrary variation of $\widetilde{\epsilon}_{yy}$ does not modify the local metric, $\widetilde{\epsilon}_{yy}$ can be obtained directly from the minimization of $\mathcal{F}_S$. Then $\widetilde{\epsilon}_{yy}=- \frac{1}{2}\widetilde{\epsilon}_{xx}$, $\widetilde{\sigma}_{yy}=0$ and $\widetilde{\sigma}_{xx}=\frac{3\eta }{4} \widetilde{\epsilon}_{xx}$. These relationships define a generalized Poisson effect, and $\widetilde{\epsilon}_{yy}=-\nu \widetilde{\epsilon}_{xx}$. 

Denoting the Gaussian curvature as $\kappa_G= \kappa_1 \kappa_2 $, we have
\begin{equation}
\widetilde{\epsilon}_{xx} \approx \frac{\kappa_{G}}{2} y^2 - \epsilon 
\label{ResidualStrainX} \text{.}
\end{equation} 
For $\widetilde{\epsilon}_{yy}=- \frac{1}{2} \widetilde{\epsilon}_{xx}$, the stretching energy (per unit length) stored in the cross-section is $\mathcal{U}_S=\frac{3 \eta }{8} \int_{-\frac{W}{2}}^{\frac{W}{2}} \widetilde{\epsilon}_{xx}^{\ 2} \ dy$. Then,
\begin{equation}
\mathcal{U}_S = \frac{3 \eta}{4} \left[ \frac{\kappa_G^{\ 2}}{20} \left( \frac{W}{2} \right)^5 - \frac{\kappa_G}{3} \left( \frac{W}{2} \right)^3 \epsilon + \left( \frac{W}{2} \right) \epsilon^2 \right]  
\label{StretchingEnergy_perLength} \text{.}
\end{equation} 
Minimization of Eq.\ref{StretchingEnergy_perLength} leads to $\epsilon=\frac{1}{24} \kappa_G W^2 $, therefore
\begin{equation}
\widetilde{\sigma}_{xx} = \frac{3 \eta }{8} \kappa_G \left( y^2 - \frac{1}{12} W^2 \right)  
\label{ResidualStrainX2} \text{.}
\end{equation}
When $\kappa_G=\kappa_0^{\ 2}$ this equation provides the residual stress used to determine anisotropic wrinkles in the main text. 
Moreover, Eq.\ref{ResidualStrainX2} changes sign at the neutral lines $y=\pm \frac{1}{\sqrt{3}}\frac{W}{2}$. For $\kappa_G > 0$, the stress is compressive at the centre of the strip ($-\frac{1}{\sqrt{3}}\frac{W}{2} < y < \frac{1}{\sqrt{3}}\frac{W}{2}$), while for $\kappa_G <  0$ the compression is at the edges ($\frac{1}{\sqrt{3}}\frac{W}{2} < \left| y \right| < \frac{W}{2}$). Since local compression drives buckling instabilities, the sign of the natural Gaussian curvature determines the spatial distribution of wrinkles, as demonstrated in Fig.~\ref{fig:Designs}-1.

\begin{figure}
\centering
\includegraphics[width=1.\linewidth]{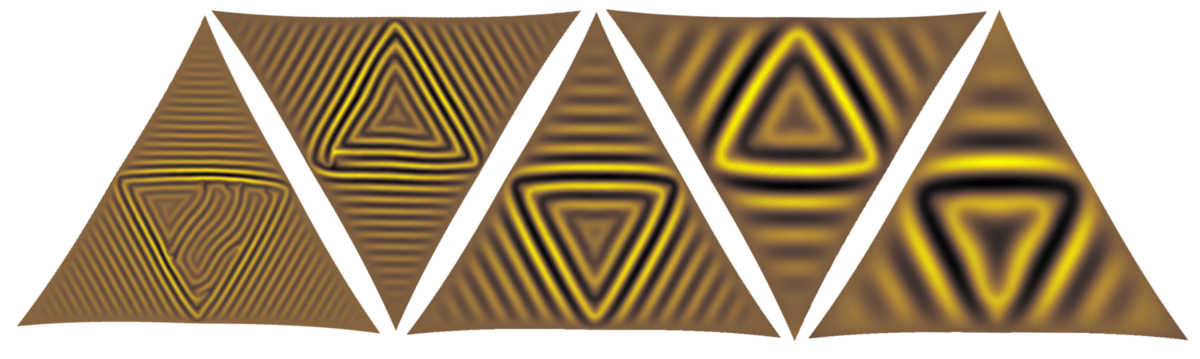}
\caption{Simulations of triangular cuts with increasing thickness (left to right). This demonstrates the effect of the variation of $\tau$ (via the physical thickness) on the wavelength.}
\label{fig:vart}
\end{figure}

So far, we calculated the stresses of planar unbuckled configurations, where the longitudinal curvature is zero but the one along the meridional lines (cross-sections) remains undetermined. The value of this curvature, denoted $\kappa_{\perp}$, does not modify the current metric and can be calculated from a simple balance between the bending and hydrostatic energies. Indeed, the bending energy per unit length is
\begin{equation}
\mathcal{U}_B = \frac{W}{2} \left[ \kappa_2^{\ 2} -  \left( \kappa_{\perp} - \kappa_1 \right) \kappa_2  + \left( \kappa_{\perp} - \kappa_1 \right)^2  \right]
\label{BendingEnergyCylinder} \text{.}
\end{equation} 
Since the cross-section profile is $h \approx \frac{1}{2} \kappa_{\perp} \left( y^2 - \frac{1}{4} W^2 \right) $ (the long edges of the shell remain at the zero level), the hydrostatic energy per unit length is given by
\begin{equation}
\mathcal{U}_H = \frac{1}{2} \int_{-\frac{W}{2}}^{\frac{W}{2}} h^2 dy=\frac{2}{15} \left( \frac{W}{2}\right)^5  \kappa_{\perp}^{\ 2}  
\label{GravitationalEnergyCylinder} \text{.}
\end{equation}
The condition $\frac{d}{d \kappa_{\perp}} \left(\mathcal{U}_B +  \mathcal{U}_H \right) =0 $ leads to $-\frac{1}{R_2} - \frac{2}{R_1} + 2 \kappa_{\perp} + \frac{4}{15} \left( \frac{W}{2} \right)^4 \kappa_{\perp}=0$. Thus, for small $W$, $\kappa_{\perp}\approx \kappa_1 + \frac{1}{2} \kappa_2$ and for spherical strips we obtain $\kappa_{\perp} \approx \frac{3}{2} \kappa_0$.

\subsection{Post-Buckling configurations of toroidal strips} \label{sec:toroidal2}

Since the stresses of the planar configuration of the toroidal strips are
\begin{equation}
\widetilde{\sigma}_{xx}=\frac{3 \eta}{8} \kappa_G \left( y^2 - \frac{1}{12} W^2  \right) \text{ ,\ } \widetilde{\sigma}_{yy}=0 \text{ ,\ }\widetilde{\sigma}_{xy}=0
\label{RibbonsResidualStreses} \text{,}
\end{equation}
the associated Airy Potential is $\widetilde{\chi}= \frac{3 \eta}{8} \kappa_G \left( \frac{1}{12} y^4 - \frac{1}{24} W^2 y^2 \right) $. According to the residual stress concept, the transformation $\chi \rightarrow \chi - \widetilde{\chi} $ allows us to use the F{\"o}ppl-von K\'arm\'an equations to determine post-buckling configurations. Then, Eq.\ref{StressBalanceAiryPotential} becomes
\begin{equation}
\Delta^2 \chi = \eta \left[ \left( \partial_{xy}^2 h \right)^2 - \left( \partial_x^2 h \right) \left( \partial_y^2 h \right) + \frac{3}{4} \kappa_G \right] \text{.}
\label{StressBalanceAiryPotential2} 
\end{equation} 
Since the excess of length is proportional to $ \left| \kappa_{G}\right|  \ll 1$, the amplitude of the post-buckling configuration is $ \sim \left| \kappa_G\right| ^{1/2}$. Using $\xi=\left| \kappa_G\right|^{1/2}$ as a small parameter, we apply a  perturbation scheme to determine the post-buckling configurations:
\begin{equation}
h= \xi h^{(1)} + \xi^2 h^{(2)} + ...  \text{ \ , \ \ } \chi= \chi^{(0)} +  \xi \chi^{(1)} +  \xi^2 \chi^{(2)} + ...
\label{PerturbationScheme2} \text{\ .}
\end{equation} 

To first order, Eq\ref{vonKarmanEquationsRectangular} can be written as
\begin{equation}
\Delta^2 h^{(1)}  + \left( 1 - \sigma_{xx}^{(0)}  \partial_x^2 - 2 \sigma_{xy}^{(0)}  \partial_{xy}^2 - \sigma_{yy}^{(0)}  \partial_y^2 \right) h^{(1)}   =0 \text{,} 
\label{vonKarmanEquationsRectangularFirstOrder} 
\end{equation} 
where $\sigma_{xx}^{(0)}=\partial^2_y \chi^{(0)}$, $\sigma_{yy}^{(0)}=\partial^2_x \chi^{(0)}$, $\sigma_{xy}^{(0)}=-\partial^2_{xy} \chi^{(0)}$ are the zero-order stress components and $\chi^{(0)}$ satisfies the biharmonic equation, $\Delta^2 \chi^{(0)}=0$. Since we are looking for wrinkling solutions of the form $h^{(1)}= f(y) \cos(qx)$, from Eq.\ref{vonKarmanEquationsRectangularFirstOrder} we obtain
\begin{equation}
\begin{split}
\frac{d^4 f}{dy^4} - \left( 2 q^2 + \sigma_{yy}^{(0)} \right) \frac{d^2 f}{dy^2} & + \left( q^4 + \sigma_{xx}^{(0)} q^2 + 1  \right) f  = \\ & = - 2 \sigma_{xy}^{(0)} q \frac{d f}{dy} \tan(qx) \text{.} 
\end{split}
\label{vonKarmanEquationsRectangularFirstOrder2} 
\end{equation} 
Taking $\sigma_{yy}^{(0)}=\sigma_{xy}^{(0)}=0$ (the non-longitudinal components of the stress remain at their pre-buckling values), the solution for $A(y)$ takes the form $f(y)= a \cos(\beta y)$. Here, $\beta$ determines the effective width of the buckling domain, $D \sim 1 / \beta $, and the longitudinal stress satisfies
\begin{equation}
\sigma_{xx}^{(0)}=-2 \beta^2 - q^2 - \frac{1+\beta^4}{q^2} \text{.}
\label{StresessParallelWrinkles(0)} 
\end{equation} 
The stable wavenumber $q$ minimizes the absolute value of this compressive stress. Since $\frac{1}{2}\frac{d}{dq} \left| \sigma_{xx}^{(0)} \right| = q - \frac{1 + \beta^4 }{q^3} $, the preferred wavenumber for $\beta \ll q $ is $q=1$ and $\sigma_{xx}^{(0)}=-2$. In general, $q^4 = 1 + \beta^4$ and $\sigma_{xx}^{(0)}=-2\beta^2 -2 \sqrt{1 + \beta^4}$.

By symmetry, $h^{(2)}=0$ and $\chi^{(1)}=0$, and with $h(x,y)=a \left| \kappa_G\right|^{1/2} \cos(\beta y) \cos(q x)$, Eq.\ref{StressBalanceAiryPotential2} becomes
\begin{equation}
\Delta^2 \chi = \frac{\eta \kappa_G}{2}\left\lbrace \frac{3}{2} - a^2 \beta^2 q^2 \left[ \cos(2 q x) + \cos(2\beta y) \right] \right\rbrace  \text{.}
\label{StressBalanceAiryPotential4} 
\end{equation} 
Then, the second order solutions are
\begin{equation}
\chi^{(2)} = \frac{\eta a^2 }{32} \left[ \frac{y^4}{a^2} - c y^2 - \frac{\beta^2}{q^2} \cos(2 q x) - \frac{q^2}{\beta^2} \cos(2 \beta y) \right] \text{,}
\label{AiryPotential4} 
\end{equation} 
where $c$ is a constant which will be determined. The associated stresses are
\begin{equation}
\begin{split}
\sigma_{xx}^{(2)} = &\frac{\eta a^2}{8} \left[ 3 \left(\frac{y}{a}\right)^2 - \frac{c}{2}  + q^2 \cos(2 \beta y) \right]   \text{,} \\
\sigma_{yy}^{(2)} = & \frac{\eta }{8} a^2 \beta^2 \cos(2 q x )  \text{ , \ } \sigma_{xy}^{(2)} = 0 \text{.}      
\end{split}
\label{StressComponentsSolution1} 
\end{equation}
In the close vicinity of the mid-line of the strip, the second order correction of the longitudinal stress is the smallest possible, since the energy at the buckling transition is minimized. Then, $c=2q^2$ and $a^2 \beta^2 q^2 = \frac{3}{2}$, so $\sigma_{xx}^{(2)} = \eta O(y^4)$. For spherical strips $\kappa_G=\kappa_0^{\ 2}$, thus using $\eta \kappa_G = \frac{4}{3} \tau $, the stresses to second order are
\begin{equation}
\sigma_{xx} = -2 \beta^2 - 2 \sqrt{1+\beta^4} + \tau O(y^4) \text{\ , \ } \sigma_{yy} = \frac{\tau}{4} \cos(2qx) \text{.}      
\label{StressComponentsSolution2} 
\end{equation}

\begin{figure}[t]
\centering
\includegraphics[width=1.0\linewidth]{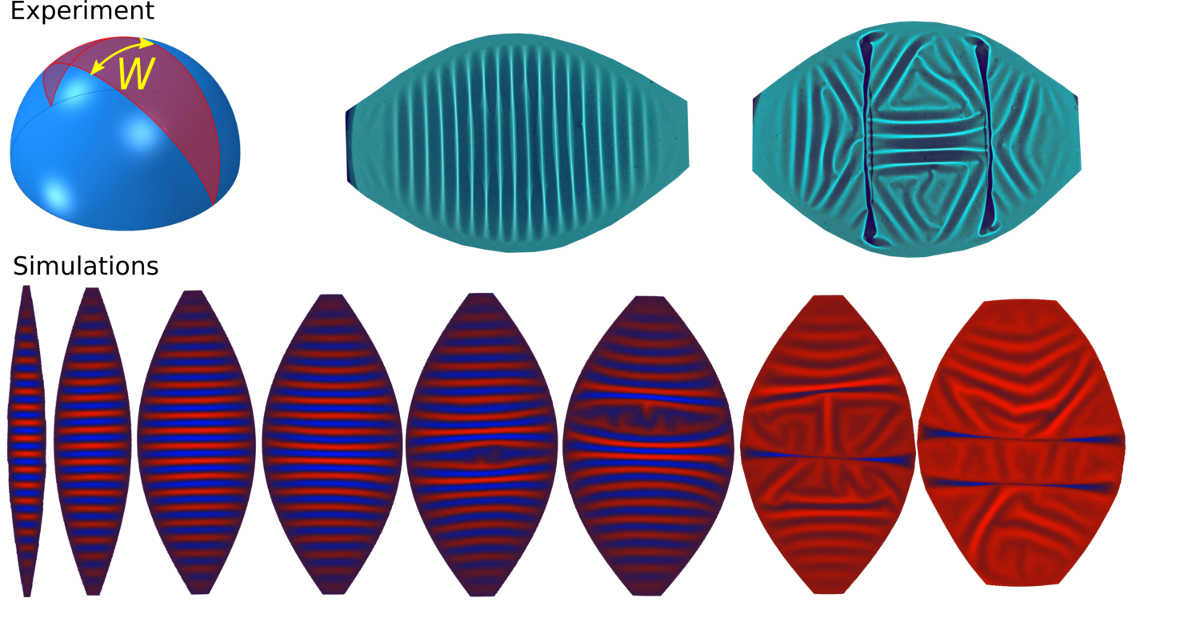}
\caption{A wrinkle to fold transition in a slice of width $W$, cut out of a spherical shell: experiments (top) and simulations (bottom).}
\label{fig:WTF-slices}
\end{figure}

\begin{figure}
\centering
\includegraphics[width=1.0\linewidth]{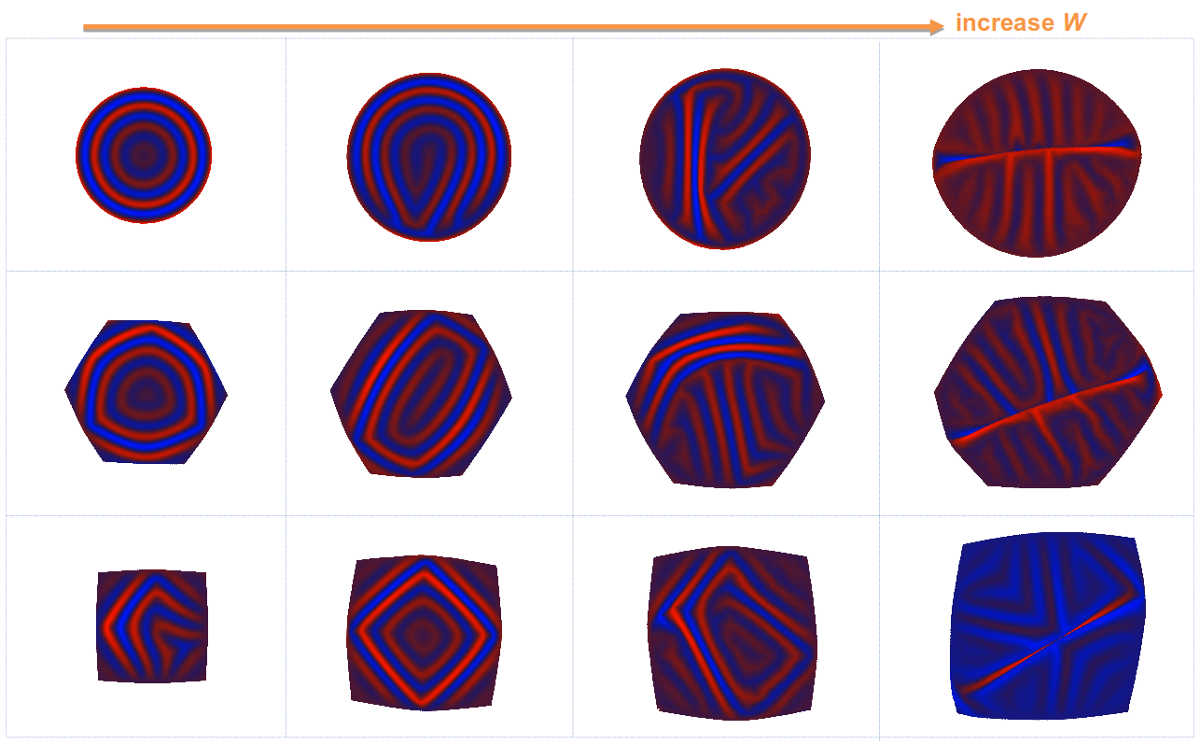}
\caption{A wrinkle to fold transition in caps, hexagons and squares with characteristic width $W$, cut out of a spherical shell -- simulations.}
\label{polygons}
\end{figure}

\end{document}